# *Time evolution of electron waves in graphene superlattices*

David E. Fernandes[a], Manuel Rodrigues[a], Gabriel Falcão[a] and

Mário G. Silveirinha[a,b] [1]

[a] *Instituto de Telecomunicações and Department of Electrical Engineering, University of Coimbra, 3030-290, Coimbra, Portugal*

[b]*Instituto Superior Técnico-University of Lisbon, Avenida Rovisco Pais, 1, 1049-001 Lisboa, Portugal*

## Abstract

The time evolution of electron waves in graphene superlattices is studied using both microscopic and "effective medium" formalisms. The numerical simulations reveal that in a wide range of physical scenarios it is possible to neglect the granularity of the superlattice and characterize the electron transport using a simple effective Hamiltonian. It is verified that as general rule the continuum approximation is rather accurate when the initial state is less localized than the characteristic spatial period of the superlattice. This property holds even when the microsocopic electric potential has a strong spatial modulation or in presence of interfaces between different superlattices. Detailed examples are given both of the time evolution of initial electronic states and of the propagation of stationary states in the context of wave scattering. The theory also confirms that electrons propagating in tailored graphene superlattices with extreme anisotropy experience virtually no diffraction.

---

[1] To whom correspondence should be addressed: E-mail: mario.silveirinha@co.it.pt

# I. Introduction

Graphene is a carbon-based material where the atoms are arranged in a honeycomb lattice [1-8]. This genuinely two-dimensional material is characterized by unusual and remarkable electronic properties, including a "relativistic"-type spectrum, such that the propagation of low-energy electrons in graphene is described by the massless Dirac equation [3]. These properties pushed graphene into the frontline of condensed matter physics research [1-7].

Interestingly, it may be feasible to gain additional control over the transport properties of electrons in graphene by artificially introducing a new length scale into the system in the form of a periodic electrostatic potential [9-16]. These structures, known as graphene superlattices, may be realized using different techniques, such as with periodically patterned gates, with the deposition of adatoms on graphene's surface, or using a crystalline substrate [17-24].

Rooted in the superlattice concept, it was recently proposed that it may be possible to extend to electronics some phenomena and devices originally discussed in the context of electromagnetism [25, 26], such as the "perfect lens" [27, 28] or an electron "wormhole" [29]. In these works, the propagation of the electrons in the superlattices was studied using an effective medium approach, wherein the granular details of the superlattices are homogenized [27]. Within this framework the structure is regarded as a continuum and the dynamics of the wave function envelope is described by an effective Hamiltonian.

The main objective of the present work is to demonstrate that the effective medium theory proposed in Ref. [27] can be used to determine the time evolution of electronic states in graphene superlattices. To this end, we develop a finite-difference time-domain (FDTD) algorithm to characterize the propagation of electron waves in superlattices using both microscopic and macroscopic formalisms. We present a detailed comparison of the physical response predicted by the two approaches. It is important to highlight that the application of numerical methods to the Schrödinger and Dirac equations is well known and has been widely reported in the open literature (e.g. Refs. [30-39]). In particular, Refs. [30, 31] study the propagation of electron waves in graphene heterojunctions using the FDTD method. However, the key novelty of our work is the verification that the effective medium formalism developed in Ref. [27] can be used to predict the physical response in the frame of *time evolution* problems as well as in the frame of *stationary state* problems in complex propagation scenarios. We are unaware of similar studies in related physical platforms. To this end, a FDTD algorithm based on a leapfrog update scheme [40] is developed and applied to the characterization of graphene superlattices using both the macroscopic and microscopic models. It is underlined that the theory developed in Refs. [30, 31] cannot be directly applied to the propagation of electron waves in the context of the effective medium model considered here, and this is the reason why we develop our own numerical scheme to solve the modified Dirac equation. It is important to make clear that what we designate here by "microscopic model" corresponds to the description of the electron wave propagation in graphene using the two-dimensional Dirac equation, which is itself an effective medium theory. Our effective Hamiltonian corresponds thus to a second level of homogenization. The Dirac equation is a valid starting point for a microscopic theory when the period of the superlattice is much larger than the atomic constant of graphene. Furthermore, the Dirac equation has been the basis of several successful theoretical predictions supported by experiments, such as the negative refraction of Dirac fermions and the connection between the optical conductivity of graphene and the fine structure constant [41, 42]. One of the nontrivial aspects of an effective medium approach is the correct formulation of boundary conditions at interfaces. Here, the boundary condition introduced in Ref. [29] is implemented in the FDTD code, and its validity is numerically confirmed in complex propagation scenarios.

The article is organized as follows. In Sec. II the effective medium model is succinctly reviewed. Section III describes the FDTD algorithm used to characterize the time evolution of electron waves. In Secs. IV and V, the theory is applied to the propagation of stationary states and to the time evolution of initial electronic states, respectively. Finally, in Sec. VI conclusions are drawn.

## II. Effective medium model

The propagation of the charge carriers in graphene may be described using the massless Dirac equation [3]:

$$i\hbar \frac{\partial}{\partial t}\boldsymbol{\psi} = \hat{H}\boldsymbol{\psi}, \tag{1}$$

being $\hat{H} = -i\hbar v_F \left( \boldsymbol{\sigma}_x \frac{\partial}{\partial x} + \boldsymbol{\sigma}_y \frac{\partial}{\partial y} \right) + V(\mathbf{r})$ the "microscopic" Hamiltonian operator near the *K* point, where $V$ is the microscopic potential, $\boldsymbol{\psi} = \{\Psi_1, \Psi_2\}$ is the two-component pseudospinor, $v_F \approx 10^6 m/s$ is the Fermi velocity and $\boldsymbol{\sigma}_x, \boldsymbol{\sigma}_y$ are the Pauli matrices. It is relevant to mention that for superlattices made from adatoms or for graphene-boron nitride superlattices the Hamiltonian gains an additional $\boldsymbol{\sigma}_z$ component [43]. Such a term can be easily incorporated into our FDTD discretization, but for simplicity in what follows we focus on superlattices described by a microscopic electrostatic potential with a one-dimensional spatial variation of the form $V(x) = V_{av} + V_{osc}\sin(2\pi x/a)$ (see Fig. 1**a**). Here, $V_{av}$ is the average potential and $V_{osc}$ is the peak amplitude of the oscillating part of the potential.

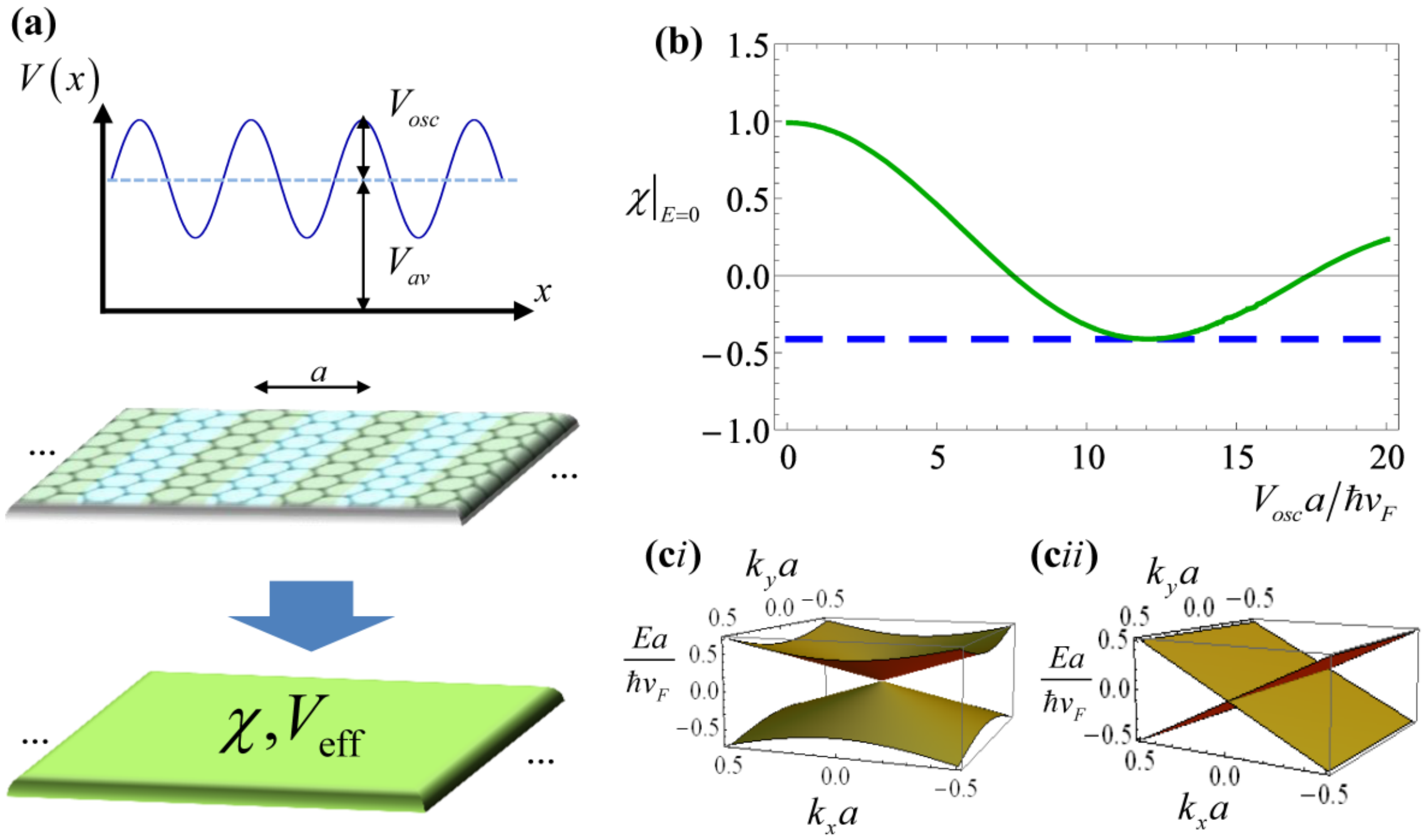

Fig. 1 (color online) **(a)** Sketch of a graphene superlattice characterized by a sinusoidal-like periodic electrostatic potential $V(x)=V_{\rm av}+V_{\rm osc}\sin(2\pi x/a)$. **(b)** Anisotropy ratio as a function of $V_{\rm osc}$. **(c)** Energy dispersion of *i*) Pristine graphene ($\chi=1$) and *ii)* Graphene superlattice characterized by an extreme anisotropy ($\chi=0$).

By solving the Dirac equation (1) it is possible to completely characterize the wave function $\boldsymbol{\psi}$ in both spatial and time domains. However, since the microscopic potential $V$ has a complex spatial dependence, this approach may be computationally demanding and provides limited physical insights.

A solution to reduce the complexity of the problem is to use effective medium methods. It was recently shown that the electronic states with the pseudo-momentum near the Dirac $K$ point can be accurately modeled using an effective medium framework [27, 29]. In this approach, the microscopic potential is homogenized and the effective Hamiltonian treats the superlattice as a continuum characterized by some effective parameters [27, 29]. For an unbounded superlattice the effective Hamiltonian is of the form:

$$\left(\hat{H}_{ef}\boldsymbol{\psi}\right)(\mathbf{r})=\left[-i\hbar v_F\boldsymbol{\sigma}(\chi)\cdot\nabla+V_{av}\right]\cdot\boldsymbol{\psi}(\mathbf{r}), \quad (2)$$

where $\boldsymbol{\sigma}(\chi)=\boldsymbol{\sigma}_x\hat{\mathbf{x}}+\chi\boldsymbol{\sigma}_y\hat{\mathbf{y}}$ and $\chi$ is an effective medium parameter designated by *anisotropy ratio*. The value of $\chi$ depends on the amplitude of the fluctuating part of the microscopic potential $V_{osc}$ and is numerically determined using the "first principles" homogenization approach described in Ref. [27]. The explicit dependence of $\chi$ on the peak amplitude of the oscillating part of the potential $V_{osc}$ is shown in Fig. 1**b)**, and reveals that the proper tuning of $V_{osc}$ may enable a regime characterized by an extreme anisotropy, where $\chi=0$. It is well known that graphene superlattices may have strongly anisotropic Dirac cones and particle velocities, and may allow for the propagation of electron waves with virtually no diffraction [18, 19, 29]. The stationary states energy dispersion obtained with the macroscopic framework is given by [27]:

$$\left|E-V_{av}\right|=\hbar v_F\sqrt{k_x^2+\chi^2k_y^2}, \quad (3)$$

where $E$ is the energy of the electrons and $\mathbf{k}=\left(k_x,k_y\right)$ is the wave vector associated with the electronic state, measured with respect to the $K$ point. In pristine graphene the anisotropy ratio is unity ($\chi=1$) and the energy dispersion is determined by the usual Dirac cone, as depicted in Fig. 1**c***i*). This linear energy dispersion corresponds to an isotropic propagation velocity. Quite differently, the energy dispersion for a superlattice with extreme anisotropy ($\chi=0$) corresponds to a Dirac cone stretched along the *y*-direction, as shown in Fig. 1**c***ii*), so that the electrons are allowed to propagate only along the *x*-direction. It is important to note that for sufficiently strong modulations of the electric potential new Dirac points can emerge in the energy spectrum [17, 21-24]. This effect is fully described by the microscopic theory [Eq. (1)] but not by the effective Hamiltonian which predicts a unique Dirac point.

The minimum value of $V_{osc}$ that leads to an extreme anisotropy is $V_{osc}a/\hbar v_F \approx 7.55$. Interestingly, this amplitude for $V_{osc}$ is *precisely* coincident with an analytical solution of Ref. [12] for new zero-energy states (due to the zero-crossing of the energy spectrum for large $k_y$) and for a strong enhancement of the conductance in superlattices with a sinusoidal profile. Since in an extreme anisotropy regime there is a strong enhancement of the electric response [44], it follows that the continuum approximation captures the essence of the findings of Ref. [12]. Furthermore, it will be shown that despite the emergence of new Dirac points the effective Hamiltonian describes extremely well the electron wave propagation in graphene superlattices with a strong potential modulation when the electron state has a characteristic width a few times larger than the lattice period.

## III. The FDTD algorithm

The knowledge of the time dynamics of electrons is essential to predict the response of graphene-based electronic devices. As a starting point, it is convenient to consider a generalized Schrödinger equation with a fictitious source term:

$$i\hbar \frac{\partial}{\partial t}\boldsymbol{\psi} = \hat{H}\boldsymbol{\psi} + i\hbar v_F \mathbf{j}. \qquad (4)$$

The term $\mathbf{j} = \begin{pmatrix} j_1 & j_2 \end{pmatrix}^T$ may be regarded as an external source that injects carriers into the system. As explained later with detail, this fictitious source is useful to characterize extended (non-localized) stationary energy states. In case of heterojunctions with a spatially varying anisotropy ratio $\chi$, the macroscopic Hamiltonian (2) should be written as:

$$\hat{H}_{ef} = -i\hbar v_F \left( \sigma_x \frac{\partial}{\partial x} + \left( \frac{1}{2}\chi \frac{\partial}{\partial y} + \frac{1}{2}\frac{\partial}{\partial y}\chi \right)\sigma_y \right) + V_{ef}(\mathbf{r}), \qquad (5a)$$

$$V_{ef}(\mathbf{r}) = V_{av}(\mathbf{r}) + \frac{\partial u}{\partial x}\hbar v_F. \qquad (5b)$$

We replaced $\chi\sigma_y i\frac{\partial}{\partial y}\rightarrow\frac{1}{2}\chi\sigma_y i\frac{\partial}{\partial y}+\frac{1}{2}\frac{\partial}{\partial y}\chi\sigma_y i$ to ensure that the Hamiltonian remains Hermitian when the anisotropy ratio $\chi$ depends explicitly on the $y$ coordinate. In addition, a spatially dependent $\chi$ requires that the macroscopic potential $V_{av}$ is transformed as $V_{av}\rightarrow V_{av}+\frac{\partial u}{\partial x}\hbar v_F$ (Eq. 5b), where $u=u\left(\chi\right)$ is defined as in Eq. (A3) of Appendix A. This transformation is required for the correct modeling of the wave propagation at an interface between distinct superlattices, as discussed in Appendix A.

For conciseness, next we present a unified description of the FDTD method that applies to both the microscopic and macroscopic (effective Hamiltonian) approaches. It is implicit that in the microscopic approach $\chi=1$ and $V_{ef}=V$. The system (4) with the Hamiltonian (5) can be spelled out as:

$$\frac{\partial\Psi_1}{\partial t}=-v_F\left(\frac{\partial}{\partial x}-i\left(\frac{\partial}{\partial y}\frac{\chi}{2}+\frac{\chi}{2}\frac{\partial}{\partial y}\right)\right)\Psi_2+\frac{V_{ef}}{i\hbar}\Psi_1+v_F j_1, \tag{6a}$$

$$\frac{\partial\Psi_2}{\partial t}=-v_F\left(\frac{\partial}{\partial x}+i\left(\frac{\partial}{\partial y}\frac{\chi}{2}+\frac{\chi}{2}\frac{\partial}{\partial y}\right)\right)\Psi_1+\frac{V_{ef}}{i\hbar}\Psi_2+v_F j_2. \tag{6b}$$

To discretize this system and obtain the time update equations in an explicit form, the spatial domain is discretized into a rectangular grid, such that the node distances along the $x$- and $y$-directions are taken equal to $\Delta_x$ and $\Delta_y$, as shown in Fig. 2. The pseudospinor is sampled at time instants separated by the time step $\Delta_t$.

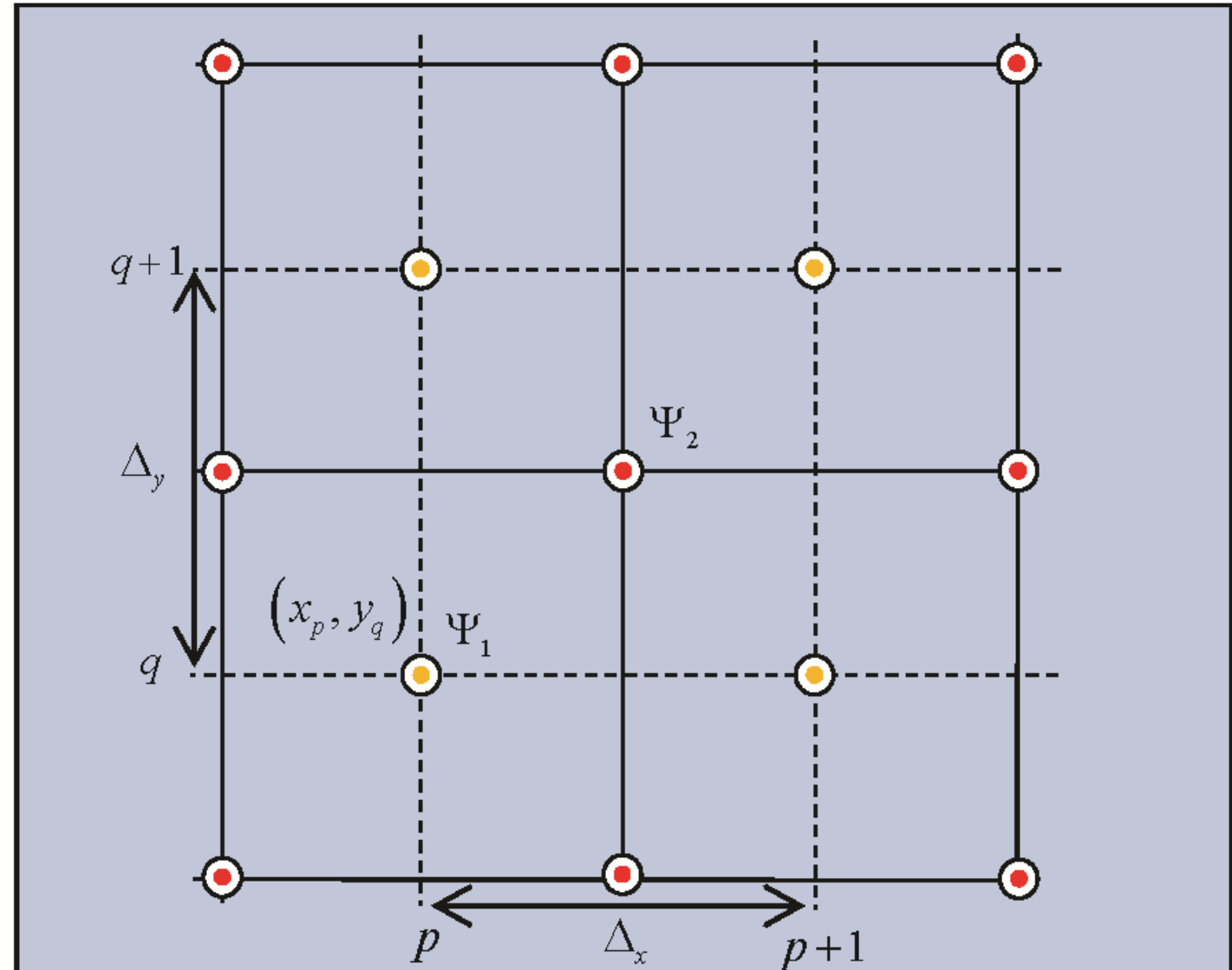

Fig. 2 (color online) The superlattice is discretized into a finite number of nodes with the pseudospinor components $\Psi_1$ and $\Psi_2$ defined in two staggered subgrids. The component $\Psi_1$ is defined over the nodes $(p,q)$ whereas $\Psi_2$ is defined over the nodes $(p+1/2, q+1/2)$ shifted by a half-grid period.

As usual, the partial derivatives are replaced by finite differences such that for a generic physical entity *F*:

$$\partial_l F(i) = \frac{F\left(i+\frac{1}{2}\right) - F\left(i-\frac{1}{2}\right)}{\Delta_l}, \tag{7}$$

where $\partial_l \equiv \partial/\partial l$ for $l = x, y, t$, and $F(x,y,t) = F\left(p\Delta_x, q\Delta_y, n\Delta_t\right) \equiv F(p,q,n)$. In our algorithm the components of the pseudospinor $\Psi_1$ and $\Psi_2$ are sampled at different points of space-time such that:

$$\Psi_1(p,q,n) \equiv \Psi_{1,p,q}^{n}, \tag{8a}$$

$$\Psi_2\left(p+\frac{1}{2}, q+\frac{1}{2}, n+\frac{1}{2}\right) \equiv \Psi_{2,p+\frac{1}{2},q+\frac{1}{2}}^{n+\frac{1}{2}}. \tag{8b}$$

Therefore, the discretization of $\Psi_1$ and $\Psi_2$ ensures that the partial derivatives of $\Psi_1$ ($\Psi_2$) in the spatial domain are defined over the same spatial subgrid as $\Psi_2$ ($\Psi_1$). The time derivatives of $\Psi_1$ and $\Psi_2$ are also defined in staggered time grids. Applying these principles to the system (6) leads to the following time update equations:

$$
\begin{aligned}
&\Psi^{n+1}_{1,p,q}\left(1-\frac{V_{p,q}}{i\hbar 2}\Delta_t\right)=\Psi^{n}_{1,p,q}\left(1+\frac{V_{p,q}}{i\hbar 2}\Delta_t\right)+v_F\Delta_t\, j^{n+\frac{1}{2}}_{1,p,q}+\\
&-v_F\Delta_t\Bigg[\left(\frac{1}{2\Delta_x}-i\frac{\chi_{p,q}+\chi_{p+\frac{1}{2},q+\frac{1}{2}}}{4\Delta_y}\right)\Psi^{n+\frac{1}{2}}_{2,p+\frac{1}{2},q+\frac{1}{2}}-\left(\frac{1}{2\Delta_x}+i\frac{\chi_{p,q}+\chi_{p-\frac{1}{2},q+\frac{1}{2}}}{4\Delta_y}\right)\Psi^{n+\frac{1}{2}}_{2,p-\frac{1}{2},q+\frac{1}{2}}\\
&+\left(\frac{1}{2\Delta_x}+i\frac{\chi_{p,q}+\chi_{p+\frac{1}{2},q-\frac{1}{2}}}{4\Delta_y}\right)\Psi^{n+\frac{1}{2}}_{2,p+\frac{1}{2},q-\frac{1}{2}}-\left(\frac{1}{2\Delta_x}-i\frac{\chi_{p,q}+\chi_{p-\frac{1}{2},q-\frac{1}{2}}}{4\Delta_y}\right)\Psi^{n+\frac{1}{2}}_{2,p-\frac{1}{2},q-\frac{1}{2}}\Bigg]
\end{aligned}
\tag{9a}
$$

$$
\begin{aligned}
&\Psi^{n+\frac{1}{2}}_{2,p+\frac{1}{2},q+\frac{1}{2}}\left(1-\frac{V_{p+\frac{1}{2},q+\frac{1}{2}}}{i\hbar 2}\Delta_t\right)=\Psi^{n-1/2}_{2,p+\frac{1}{2},q+\frac{1}{2}}\left(1+\frac{V_{p+\frac{1}{2},q+\frac{1}{2}}}{i\hbar 2}\Delta_t\right)+v_F\Delta_t\, j^{n}_{2,p+\frac{1}{2},q+\frac{1}{2}}\\
&-v_F\Delta_t\Bigg[\left(\frac{1}{2\Delta_x}+i\frac{\chi_{p+\frac{1}{2},q+\frac{1}{2}}+\chi_{p+1,q+1}}{4\Delta_y}\right)\Psi^{n}_{1,p+1,q+1}-\left(\frac{1}{2\Delta_x}-i\frac{\chi_{p+\frac{1}{2},q+\frac{1}{2}}+\chi_{p,q+1}}{4\Delta_y}\right)\Psi^{n}_{1,p,q+1}\\
&+\left(\frac{1}{2\Delta_x}-i\frac{\chi_{p+\frac{1}{2},q+\frac{1}{2}}+\chi_{p+1,q}}{4\Delta_y}\right)\Psi^{n}_{1,p+1,q}-\left(\frac{1}{2\Delta_x}+i\frac{\chi_{p+\frac{1}{2},q+\frac{1}{2}}+\chi_{p,q}}{4\Delta_y}\right)\Psi^{n}_{1,p,q}\Bigg]
\end{aligned}
\tag{9b}
$$

where $V_{p,q}=V_{ef}\left(p\Delta_x,q\Delta_y\right)$, $\chi_{p,q}=\chi\left(p\Delta_x,q\Delta_y\right)$, etc. In the derivation of the above equations, we use interpolation formulas such as $\Psi_1\left(p,q,n+\frac{1}{2}\right)\approx\frac{1}{2}\left(\Psi^{n}_{1,p,q}+\Psi^{n-1}_{1,p,q}\right)$ to evaluate the wave function at points not lying in the pertinent subgrid. By sequentially using the explicit update equations (9) in a leapfrog scheme [40], it is possible to determine $\Psi^{n}_{1,p,q}$ and $\Psi^{n+\frac{1}{2}}_{2,p+\frac{1}{2},q+\frac{1}{2}}$ at a generic instant of time *n* from the knowledge of the initial state of the system ($\Psi^{n}_{1,p,q}$ and $\Psi^{n+\frac{1}{2}}_{2,p+\frac{1}{2},q+\frac{1}{2}}$ calculated at *n*=0). This update scheme is completely analogous to the FDTD algorithm for electromagnetic waves [40]. In Appendix B, we formally demonstrate that our algorithm is stable provided the time step satisfies:

$$\Delta_t < \frac{1}{v_F} \frac{1}{\sqrt{\frac{1}{\Delta_x^2} + \frac{\chi^2}{\Delta_y^2}}}. \tag{10}$$

Moreover, in Appendix C, it is shown that unbounded regions (with the graphene sheet infinitely extended) can be numerically emulated using a "perfectly matched layer" (PML). For complex heterostructures the FDTD algorithm may require substantial computational resources. Given the processing power provided by current graphics processing units (GPUs), the algorithm was implemented based on parallel computing methods. A validation of the FDTD algorithm for electron waves propagating in simple graphene heterostructures is reported in the Supplementary Materials [45]. In the next sections, the algorithm is applied to graphene superlattices.

## IV. Stationary States in Graphene Superlattices

In the following, we investigate the propagation of extended stationary states (with a definite energy) in graphene superlattices using both the microscopic and effective medium formalisms.

First, we consider the propagation problem in an unbounded graphene superlattice. Within the continuum approximation described in Sect. II, the wave packet (group) velocity is given by [see Eq. (3)]:

$$\mathbf{v} = \frac{1}{\hbar} \nabla_{\mathbf{k}} E = \operatorname{sgn}\left(E - V_{av}\right) v_F \frac{1}{\sqrt{k_x^2 + \chi^2 k_y^2}} \left(k_x, \chi^2 k_y\right). \tag{11}$$

Thus, unlike in pristine graphene ( $\chi = 1$ ) in general the group velocity is not parallel to the quasi-momentum $\mathbf{k}$. In particular, in the extreme anisotropy limit ( $\chi = 0$ ) the group velocity satisfies $\mathbf{v} = \pm v_F \hat{\mathbf{x}}$, and hence in this case all the electron states flow along the $x$-direction and the superlattice supports diffractionless propagation [11-16]. To illustrate this effect, we consider the propagation of a Gaussian electron wave with initial beamwidth radius $R_G = 14.14a$ and normalized energy $E_0 a/\hbar v_F = 0.2$ in an unbounded graphene superlattice. The superlattice is characterized by $\chi = 0$, $(V_{\text{av}} - E_0) a/\hbar v_F = 0.1$, being $a = 10\text{nm}$ the lattice period. In a first stage, the superlattice is treated as a continuum.

An (extended) stationary state can be characterized with the FDTD method using a fictitious "electron source", i.e. with a suitable $\mathbf{j}$ in Eq. (4). The role of the source is to imitate the continuous flow of the incoming stationary wave packet. The explicit formulas that are used to generate an incoming Gaussian wave packet with energy $E = E_0$ are given in Appendix D. The time dependence of the fictitious source is of the form $e^{-i\omega_0 t}$ with $\omega_0 = E_0 / \hbar$, and the source is turned on at time $t = 0$ with an initial state $\boldsymbol{\psi}_{t=0} = 0$. In order to imitate an unbounded structure the computational domain is truncated with a PML. After a sufficiently long time, the wave function will reach a steady state such that the time variation of $\boldsymbol{\psi}$ is also of the form $e^{-i\omega_0 t}$ in all points of space and the integral $\int |\boldsymbol{\psi}|^2 \, dxdy$ becomes time independent. In all the calculations of the article, we used a time step $\Delta_t = 0.35 \dfrac{\Delta}{v_F \sqrt{2}} \approx 0.62\text{as}$ for a spacing between nodes $\Delta = \Delta_x = \Delta_y = 0.25\text{nm}$, consistent with the stability criterion defined in Eq. (10).

In the present example, the stationary regime in the FDTD method is reached after $16 \times 10^3$ time steps so that the propagation time is $t \approx 0.99\text{ps}$. The spatial distribution of the probability density function is depicted in Fig. 3**a)** and reveals that the beamwidth of the electron wave is unaffected by the propagation in the superlattice. Note that in our plots the *x*- (stratification) direction is the vertical axis. This confirms that within the continuum approximation a superlattice with $\chi = 0$ is insensitive to diffraction effects.

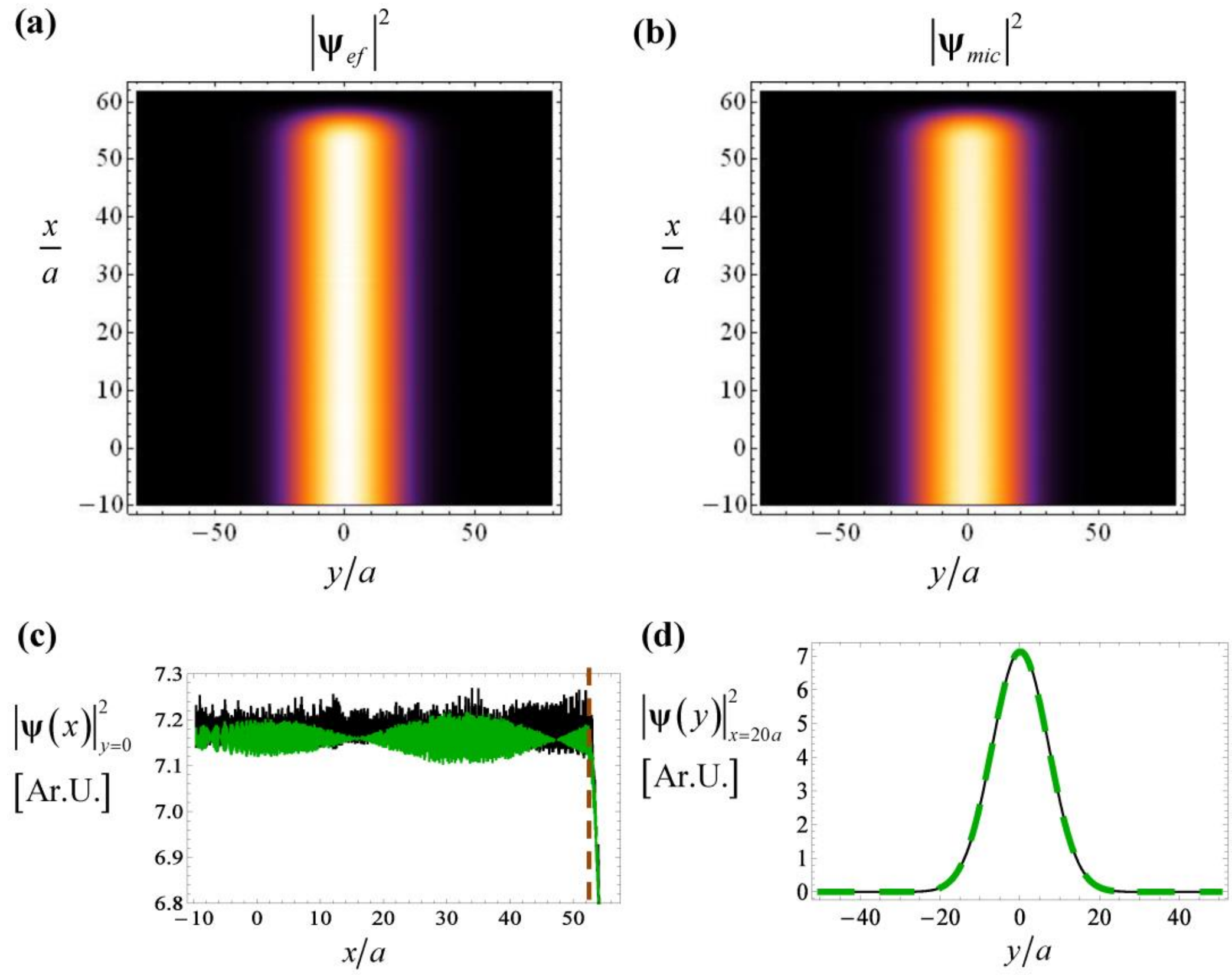

Fig. 3 (color online) **(a)** Density plot of $|\psi|^2$ calculated with the continuum approximation. **(b)** Similar to **(a)** but calculated with the exact microscopic theory. **(c)** Longitudinal profiles of the probability density function (normalized to arbitrary units) calculated using the microscopic model (black curve), and with the continuum approximation (green – light gray – curve). **(d)** Transverse profiles of the probability density function at $x = 20a$ calculated with the microscopic model (thick black curves) and with the continuum approximation (dashed green curves). In all the examples, $E_0 a/\hbar v_F = 0.2$ and $(V_{av} - E_0) a/\hbar v_F = 0.1$. The incident Gaussian electron wave has $R_G = 14.14a$ and propagates in a superlattice characterized by the anisotropy ratio $\chi = 0$ (in the microscopic model $V_{osc} a/\hbar v_F \approx 7.55$ ).

To validate these results we applied the FDTD algorithm to the corresponding microscopic structure taking into account that the microscopic potential has a sinusoidal-type spatial variation (see Fig. 1**a**)) with $V_{osc} a / \hbar v_F \approx 7.55$. This value of $V_{osc}$ corresponds to a vanishing $\chi$ in the continuum model (see Fig. 1**b**)) [27]. The probability density function obtained with the microscopic approach is depicted in Fig. 3**b**), and agrees extremely well with the effective medium theory results. This is corroborated by Figs. 3**c)** and 3**d)**, which show the longitudinal and transverse profiles of the probability density function determined using the continuum and the microscopic models. Moreover, in Figs. 4**a)**-**b)** we also represent the amplitude and phase of the two components of the pseudospinor $\boldsymbol{\psi} = \{\Psi_1, \Psi_2\}$ along the $y = 0$ line. As is well known, each component of the pseudospinor is associated with a different sublattice of graphene [3, 46]. Thus, the FDTD results show that the continuum model accurately characterizes the effective response of both sublattices. It should be noted that the region $x > 52a$ of the computational domain corresponds to the PML region, whose boundary is marked by the dashed vertical line in Fig. 3**c)** and in Fig. 4. As seen, the PML region effectively "absorbs" the electron wave without reflections, mimicking an open boundary. We numerically verified (not shown) that the results are virtually unchanged for a graphene strip with a finite transverse width *W* along the *y*-direction, being *W* a few times larger than $R_G$. Thus, in the extreme anisotropy limit the edges play no role on the electron wave propagation.

It is interesting to note that due to the granularity of the superlattice the microscopic model results have considerable fluctuations on the scale of *a*, particularly the phase of the pseudospinor oscillates wildly in each period as seen in Figs. 4**a)**-**b)**. To filter out these oscillations, the microscopic results are spatially averaged, so that for a physical entity *F* we calculate $F_{av}(x, y) = \frac{1}{a} \int_{-a/2}^{a/2} F(x + x', y) dx'$. The transverse profiles of the amplitude and phase of the pseudospinor $\boldsymbol{\psi} = \{\Psi_1, \Psi_2\}$ calculated using the effective medium theory and the microscopic model with spatial averaging are shown in Fig. 4**c)**-**d).** The results demonstrate that with the spatial averaging the microscopic and the effective medium model results are virtually coincident, which is consistent with the theory of Ref. [27]. In the rest of the article, the spatial averaging is applied to all the microscopic model results.

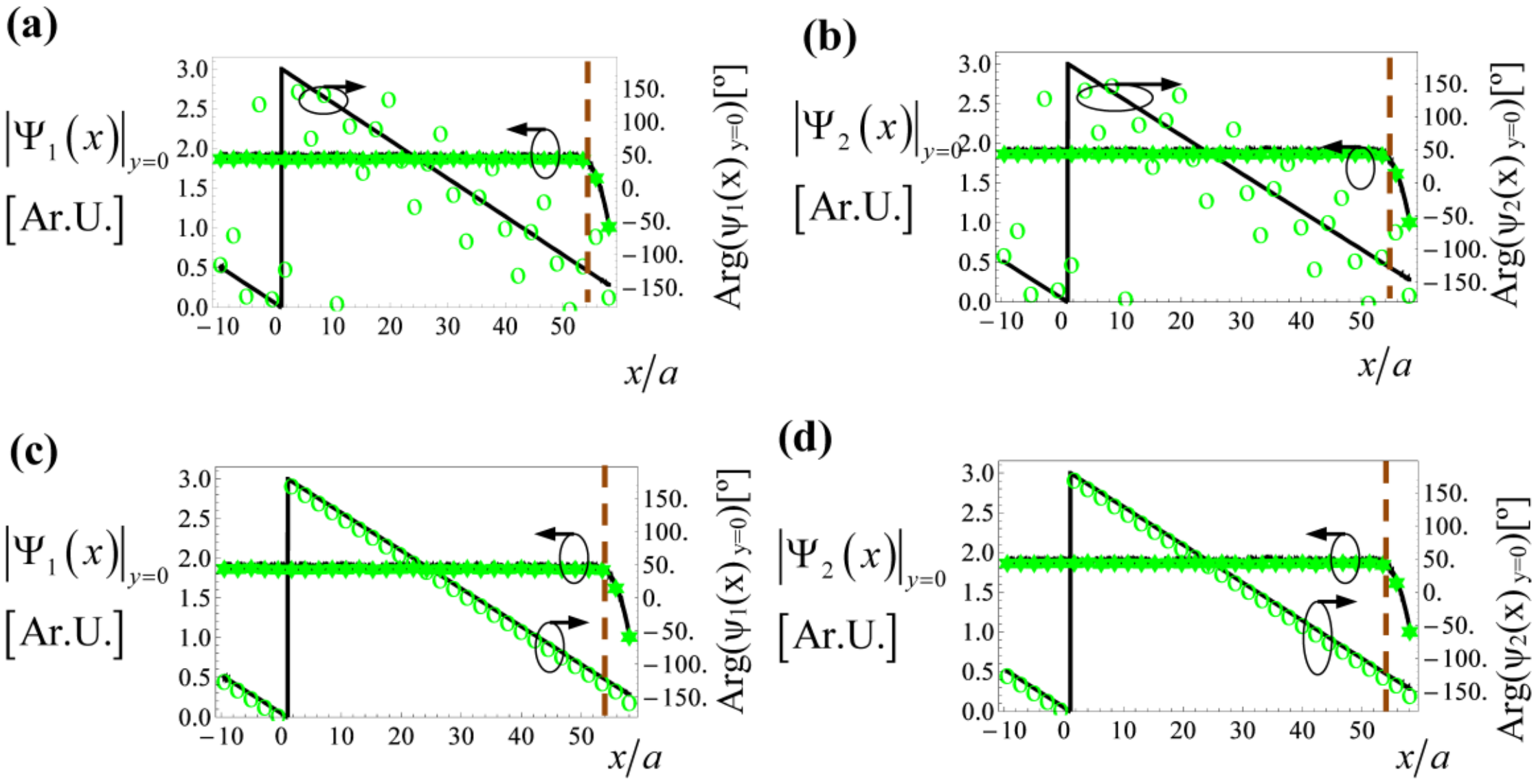

Fig. 4 (color online) **(a)** Amplitude and phase of $\Psi_1$ (along the $y=0$ line) calculated using the microscopic model (green discrete symbols) and with the continuum model (black solid curves). **(b)** Similar to **(a)** but for $\Psi_2$. **(c)** and **(d)** are similar to **(a)** and **(b)** but the microscopic results are spatially averaged. The structural parameters and the energy value are as in Fig. 3.

To determine the limits of validity of the continuum approximation, we calculated the stationary regime results for Gaussian wave packets with increasingly small radial width $R_G$ (Fig. 5). The simulations show that for $R_G = 1.9a$ the continuum model still captures the physical response of the superlattice, but for smaller beamwidths (e.g. $R_G = 0.4a$) it gives results diverging from the microscopic theory. Indeed, for $R_G = 0.4a$ the wave is diffracted by the structure leading to a significant broadening of the wave packet, as illustrated in Fig. 5**f)** . This finding is consistent with the general limits of validity of effective medium methods which are expected to break down when the wave becomes more localized than the lattice period [27]. Indeed, when $R_G < a$, electronic states with large values of the wave vector (e.g. states associated with new

Dirac points) can influence the wave propagation. These states are not described by the continuum model.

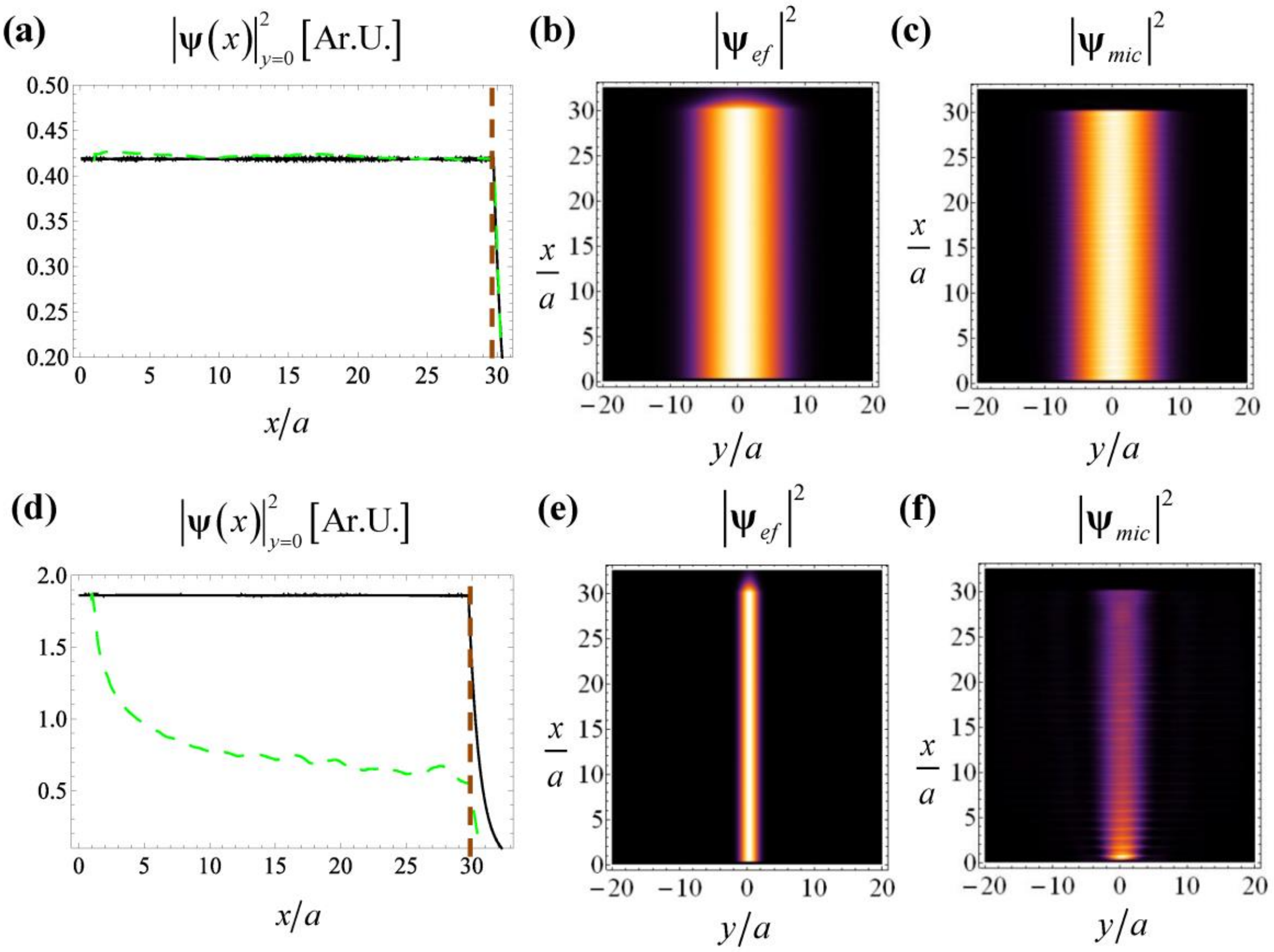

Fig. 5 (color online) Gaussian electron wave with $R_G = 1.9a$ and energy $E_0 a/\hbar v_F = 0.2$ (stationary regime). **(a)** Longitudinal profiles of $|\psi|^2$ (normalized to arbitrary units) calculated using the microscopic model (green – light gray – dashed curve), and with the continuum model (black curve). **(b)** Density plot of $|\psi|^2$ calculated with the continuum approximation. **(c)** Similar to **(b)** but calculated with the exact microscopic theory. **(d)**-**(f)** Similar to **(a)**-**(c)** but for a wave with $R_G = 0.4a$. In all the examples the structural parameters of the superlattice are as in Fig. 3.

Next, we study the wave propagation in complex graphene heterostructures. Specifically, in a first step we investigate the electron wave scattering by a superlattice nanostrip encapsulated in pristine graphene. The superlattice nanostrip has the same structural parameters as in Fig. 3 and thickness $W = 20a$. Figure 6 shows the transmission coefficient $T$ for the pseudo-spinor amplitude calculated with the FDTD code for $E = E_0$, with $E_0 a/\hbar v_F = 0.2$, as a function of the pseudo-momentum $k_y$. These results are compared with "exact" analytical results for plane wave incidence obtained using the theory of Ref. [29]. Note that the analytical results depend if the superlattice is regarded as a granular structure or as a continuum [29].

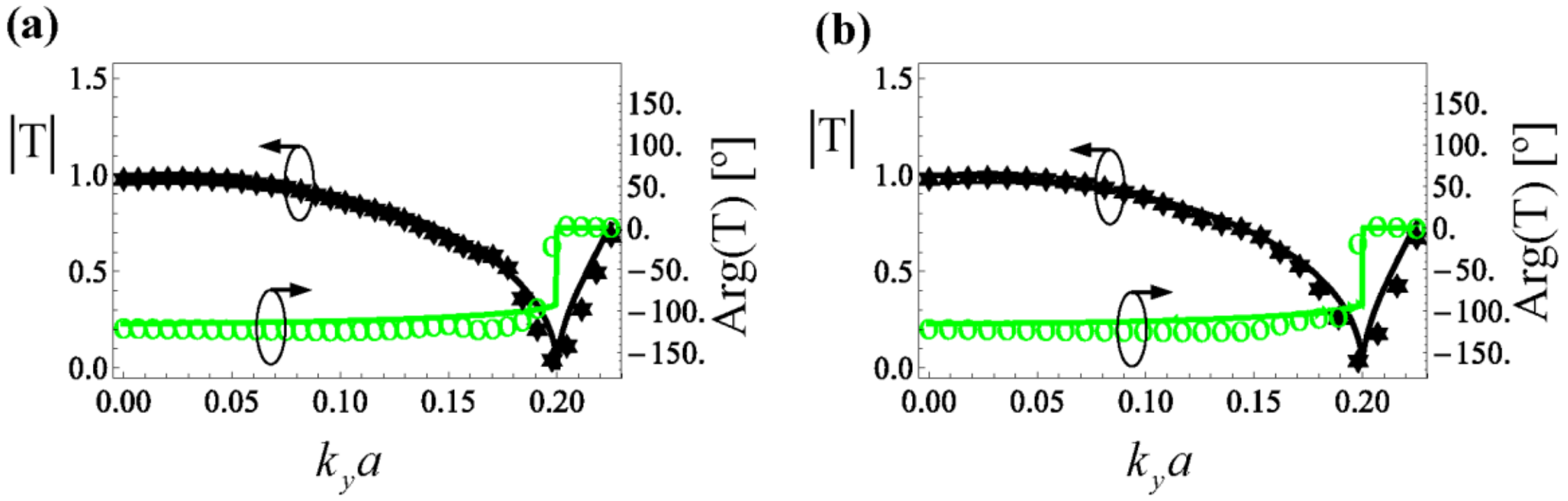

Fig. 6 (color online) **(a)** Transmission coefficient as a function of the normalized transverse quasi-momentum $k_y$ (continuum approximation). The incident electron wave propagates in pristine graphene with energy $E_0 a/\hbar v_F = 0.2$ and impinges on a superlattice nanostrip with thickness $W = 20a$ and with the same structural parameters as in Fig. 3. (**b**) is similar to **(a)** but calculated for the associated microscopic structure. In both the examples, the discrete symbols represent the results calculated using the FDTD algorithm and the solid thick curves represent the analytical results obtained using the theory of Ref. [29].

As seen in Fig. **6**, there is an excellent agreement between the FDTD results and the analytical theory for both the microscopic (Fig. **6b**)) and the continuum formulations (Fig. **6a**)). Note that as $k_y a$ increases from 0 to 0.2 the incidence angle varies from $0^o$ to $90^o$ (grazing incidence). For $k_y a > 0.2$ the incident wave is evanescent and the *x*-component of the incident wave vector becomes imaginary so that the states are not normalizable.

A crucial aspect is that within the macroscopic framework the wave function is not continuous at the interface, but it rather satisfies a boundary condition derived in Ref. [29] and further discussed in Appendix A. In the FDTD algorithm the effect of the nontrivial boundary condition is described by the function $u(\chi)$ in Eq. (5b). When $u(\chi)$ is taken equal to zero the boundary condition reduces to the continuity of the pseudospinor.

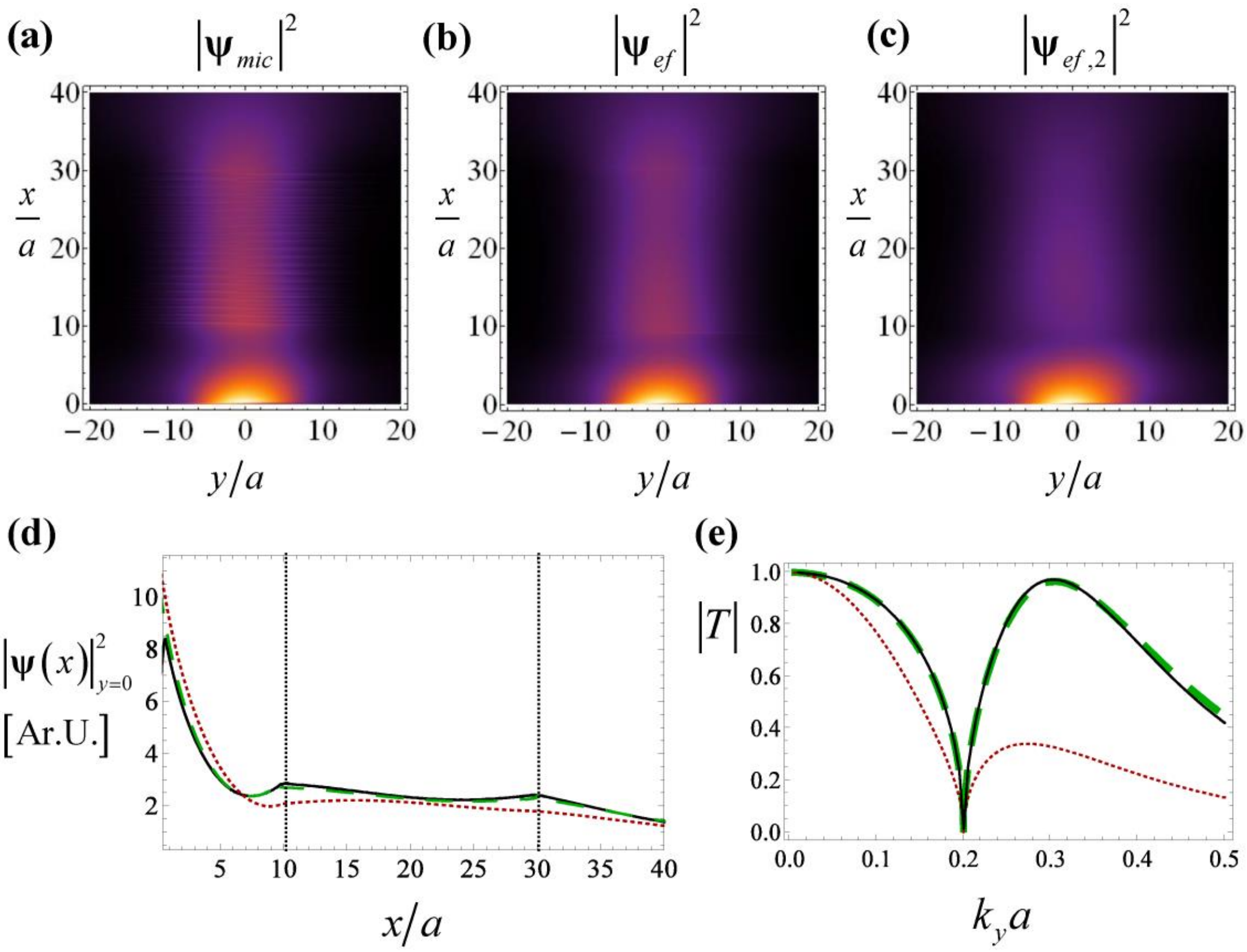

Fig. 7 (color online) An incident Gaussian electron wave with $R_G = 5a$ propagating in pristine graphene impinges on a superlattice nanostrip characterized by the anisotropy ratio $\chi = 0.25$ (in the microscopic model $V_{osc}a/\hbar v_F \approx 6.13$ ) and $(V_{av} - E_0)a/\hbar v_F = 0.1$. The energy is $E_0 a/\hbar v_F = 0.2$ and the incident angle is $\theta_i = 53°$. **(a)** Density plot of $|\psi|^2$ calculated with the exact microscopic theory. **(b)** Similar to **(a)** but calculated with the continuum model. **(c)** Similar to **(b)** but calculated with the incorrect boundary condition. **(d)** Longitudinal profiles of $|\psi|^2$ calculated using the microscopic model (black curve), with the continuum model (green – light gray – dashed curve), and with the continuum model with the incorrect boundary condition (red – dark gray – dotted curve). **(e)** Amplitude of the transmission coefficient as a function of the normalized transverse quasi-momentum $k_y$. The legend is as in **(d).**

To highlight the importance of using the correct boundary condition in the continuum approximation, we consider the scenario wherein a Gaussian electron wave with $R_G = 5a$, normalized energy $E_0 a/\hbar v_F = 0.2$ impinges on a superlattice nanostrip with an incidence angle $\theta_i = 53°$. The input interface is at $x = 10a$. The Gaussian wave is created by a fictitious source placed at the $x = 0$ plane. The graphene nanostrip is a superlattice with thickness $W = 20a$, average potential $\left(V_{\text{av}} - E_0\right)a/\hbar v_F = 0.1$ and anisotropy ratio $\chi = 0.25$ (in the microscopic model $V_{\text{osc}}a/\hbar v_F \approx 6.13$). Figure 7 shows the calculated probability density function for the microscopic theory, the effective medium theory with the correct boundary condition, and the effective medium theory assuming the continuity of the pseudospinor. The last case corresponds to setting $u = 0$ in the update equations.

As seen in Fig. 7**a)**-**c)** the continuum model only concurs with the exact microscopic theory when the correct parameter $u$ is used. This is made clear in Fig. 7**d)**, which shows the longitudinal profiles of the probability density function. The red dotted curve, which corresponds to the incorrect boundary condition (i.e. to the continuity of the macroscopic pseudospinor), underestimates the value of the wave function inside the superlattice. To have a clearer idea of the discrepancy introduced by the incorrect boundary condition we show in Fig. 7**e)** the transmission coefficient for a plane wave that impinges on the same graphene nanostrip [29]. Consistent with the results of Fig. 7**d)**, the incorrect boundary condition strongly underestimates the transmission across the nanostrip, especially for $k_y > E_0/\hbar v_F$.

## V. Time evolution of Electronic States

In the following, the problem of time evolution of a given initial electronic state is considered. This problem involves solving the FDTD equations subject to a given initial condition $\boldsymbol{\psi}|_{t=0}$. Of course, in this context one does not need to consider a fictitious source and hence we set $\mathbf{j}=0$ in Eq. (4). In this work, the initial state is assumed to be of the form (the normalization of the wave function is arbitrary):

$$\boldsymbol{\psi}(x,y,t=0)=\begin{pmatrix} 1 \\ \dfrac{\hbar v_F}{E_0-V}\left(k_{x0}+i\chi\,k_{y0}\right)\end{pmatrix} e^{i\left(k_{x0}x+k_{y0}y\right)} e^{-\frac{(x-x_0)^2+(y-y_0)^2}{R_G^2}}. \tag{12}$$

This initial state corresponds to a Gaussian wave packet initially centered at $(x_0, y_0)$ and with a radial width $R_G$. The parameters $E_0$ and $\mathbf{k}_0=\left(k_{x0},k_{y0}\right)$ (with $\left|E_0-V\right|=\hbar v_F\sqrt{k_{x0}^2+\chi^2 k_{y0}^2}$ ) play the role of "energy" and quasi-momentum of the wave packet, even though, strictly speaking the considered state is not an eigenstate of the energy operator. The vector $\mathbf{k}_0$ determines the direction of propagation. Similar to the previous sections, the computational domain is surrounded by a PML to avoid reflections from the sidewalls. Thus, after a sufficiently long time the initial state will be absorbed by the PML and the probability of finding the electron inside the computational domain approaches zero.

In the first example, the initial electronic state propagates in an unbounded superlattice characterized by an anisotropy ratio $\chi_0$ and average potential $V_0=0$, as shown in Fig. 8**a**). The parameters that characterize the initial state are $R_G=2.82a$, $\dfrac{E_0 a}{\hbar v_F}=1.9$ and $\mathbf{k}_0=(1.9,0)/a$, with $a=10\text{nm}$; the initial wave packet is centered at $(x_0, y_0)=(-15a,0)$.

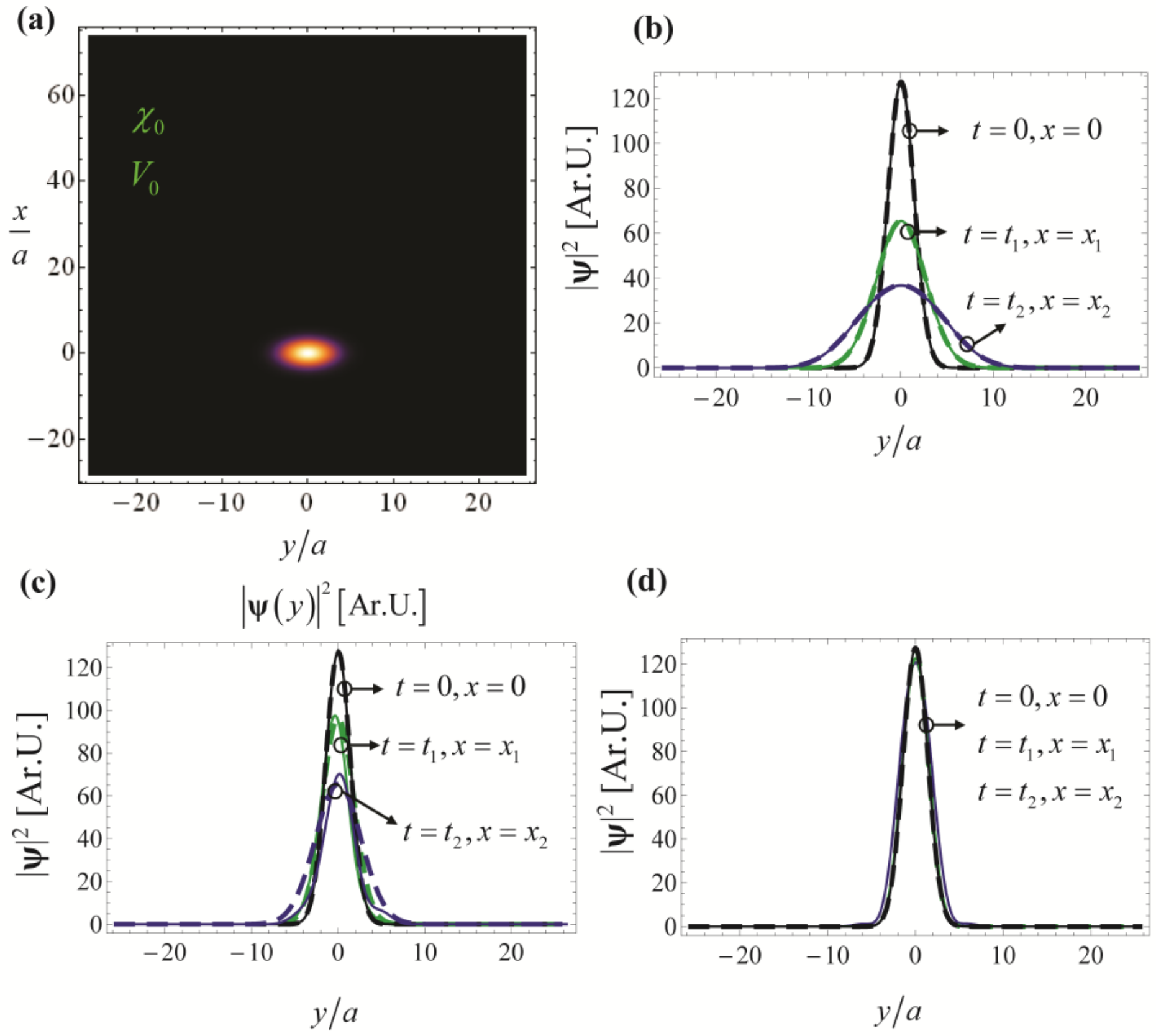

Fig. 8 (color online) **(a)** Geometry of the unbounded superlattice under study. **(b)** Transverse profile of the probability density function at $x=0$ sampled at the time instant $t=0$ (black curves), at $x_1=12.35a$ sampled at $t=t_1=2000\Delta_t$ (green – light gray – curves) and for $x_2=2x_1$ sampled at $t=t_2=4000\Delta_t$ (blue – dark gray – curves), for pristine graphene ($\chi_0=1$) with $V_0=0$. **(c)** Similar to **(b)** but for a superlattice with $\chi_0=0.7$ ($V_{osc}a/\hbar v_F\approx 3.58$). **(d)** Similar to **(b)** but for a superlattice with $\chi_0=0$ ($V_{osc}a/\hbar v_F\approx 7.55$). In all plots, the dashed lines represent the microscopic theory results, and the solid thick lines represent the continuum approximation results. The time animations of all plots are available online (multimedia view) [45].

First, we consider the electron wave propagation in pristine graphene ($\chi_0=1$). Using the FDTD algorithm the wave function was sampled at three different time instants, $t=0$, $t=t_1=2000\Delta_t$ and $t=t_2=4000\Delta_t$, with the time step $\Delta_t$ defined as in Sect. IV. Figure 8**b)** represents the transverse profiles of probability density function at the *x=const.* lines wherein the amplitude of $|\psi(x,t)|^2_{y=0}$ is maximal for a fixed *t*. As expected, the time evolution of the initial electronic state causes the Gaussian wave packet to broaden and increase its characteristic size. Indeed, in pristine graphene there is no preferred direction of motion because the group velocity is independent of the direction of propagation, and hence the wave is diffracted. Note that in this example the microscopic and continuum results are coincident because there is no microscopic potential.

We did similar studies for the case of superlattices characterized by anisotropy ratios $\chi_0 = 0.7$ and $\chi_0 = 0$ ($V_{osc} a / \hbar v_F \approx 3.58$ and $V_{osc} a / \hbar v_F \approx 7.55$, respectively, in the microscopic model). The corresponding results are depicted in Figs. 8**c**) and 8**d)**, respectively, and the associated time animations can be found in Ref. [45]. As seen, consistent with the findings of Sect. IV, as the value of $\chi_0$ approaches zero the broadening of the wave packet becomes insignificant. In particular, for $\chi_0 = 0$ (Fig. 8**d)**) the electronic state is unaffected by diffraction and the shape of the wave front does not change with time. Importantly, the results obtained with the exact microscopic model are largely coincident with the results obtained with our effective medium theory, confirming that the effective Hamiltonian can be used − when the initial state is less localized than the period $a$ − to predict the time evolution of the electronic states in a superlattice [27].

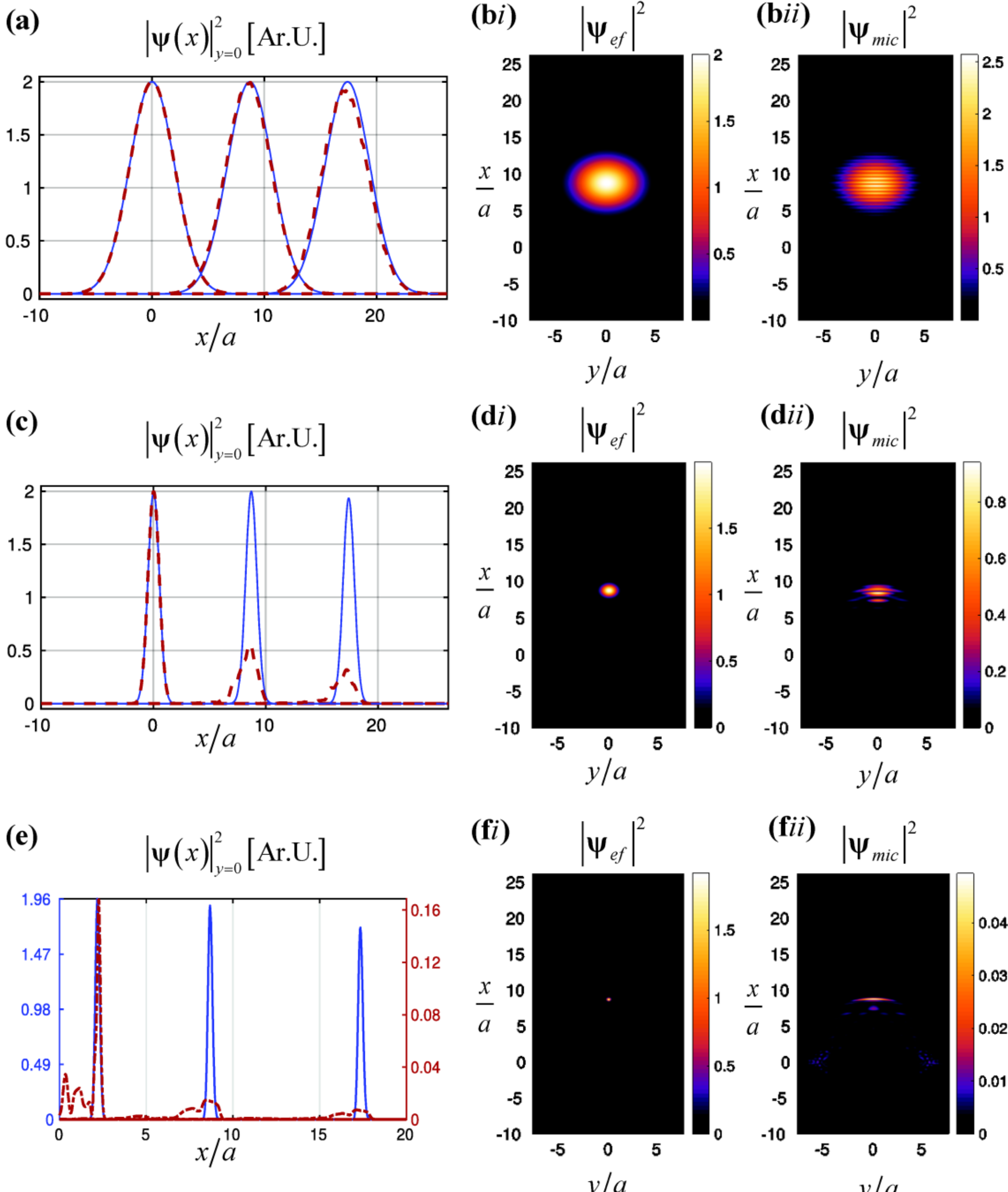

Fig. 9 (color online) **(a)** and **(b)** Time evolution of an initial electronic state with $R_G = 4a$ **(a)** Longitudinal profiles of the probability density function at $x = 0$ sampled at the time instants $t = 0$, $t = t_1 = 1000\Delta_t$ and $t = t_2 = 2000\Delta_t$. Solid blue curves: continuum approximation. Dashed red curves: microscopic theory. **(b)** Spatial distribution of the probability density function calculated using the (*i*) continuum model and (*ii*) the microscopic model at the time instant $t = 1000\Delta_t$. **(c)** and **(d)** are similar to **(a)** and **(b)** but calculated for an initial electronic state with $R_G = 1a$. **(e)** and **(f)** are similar to **(a)** and **(b)** but calculated at $t = t_1 = 250\Delta_t$, $t = t_2 = 1000\Delta_t$ and $t = t_3 = 2000\Delta_t$ for an initial electronic state with $R_G = 0.25a$. In all the examples the superlattice is characterized by an anisotropy ratio $\chi_0 = 0$ ($V_{osc}a/\hbar v_F \approx 7.55$).

Similar to the previous section, the continuum model breaks down when the characteristic size of the initial electronic state is comparable to the period of the superlattice. This property is made clear by the results of Fig. **9**, which shows that the effective medium theory still concurs extremely well with the microscopic theory for an initial state with $R_G = 4a$ (Figs. **9a)** and **9b**)). However, when the width of the initial state is decreased to $R_G = a$ or $R_G = 0.25a$ (Figs. 9**c)**-9**f)**) the spatial spectrum of the initial state is formed by harmonics with large wave vectors that are not described by the continuum approximation, and hence the two formulations predict quite different time evolutions.

In the final example, we consider an electron wave propagating in a cascade of several superlattices embedded in pristine graphene ($\chi_0 = 1$ and $V_0 = 0$) (Fig. 10**a)**). With reference to Fig. 10**a)**, the parameters of the superlattices are: anisotropy ratio $\chi_{1,2} = \mp 0.2$ ($V_{osc,1}a/\hbar v_F \approx 8.90$ and $V_{osc,2}a/\hbar v_F \approx 6.41$ in the microscopic model), average potential $V_{1,2}a/\hbar v_F = \mp 0.5$ and thicknesses $W_1 = 7.5a$ and $W_2 = 15a$. The initial state propagates along the *x*-direction and has a radial width $R_G = 2.82a$ and $E_0 a/\hbar v_F = 1.9$. Similar to the previous example, the time evolution problem is solved using the FDTD algorithm and the longitudinal profile of probability density distribution is sampled at specific time instants: $t = 0$, $t_1 = 2000\Delta_t$, $t_2 = 3800\Delta_t$, $t_3 = 5600\Delta_t$ and $t_4 = 7600\Delta_t$. As depicted in Fig. 10**b)**, the sampling times are such that the maximum of $|\psi|^2$ lies at the mid-plane of each superlattice nanostrip in Fig. 10**a)**. Remarkably, despite the complexity of the microscopic potential, the results obtained with the continuum model are nearly overlapped with the microscopic theory results.

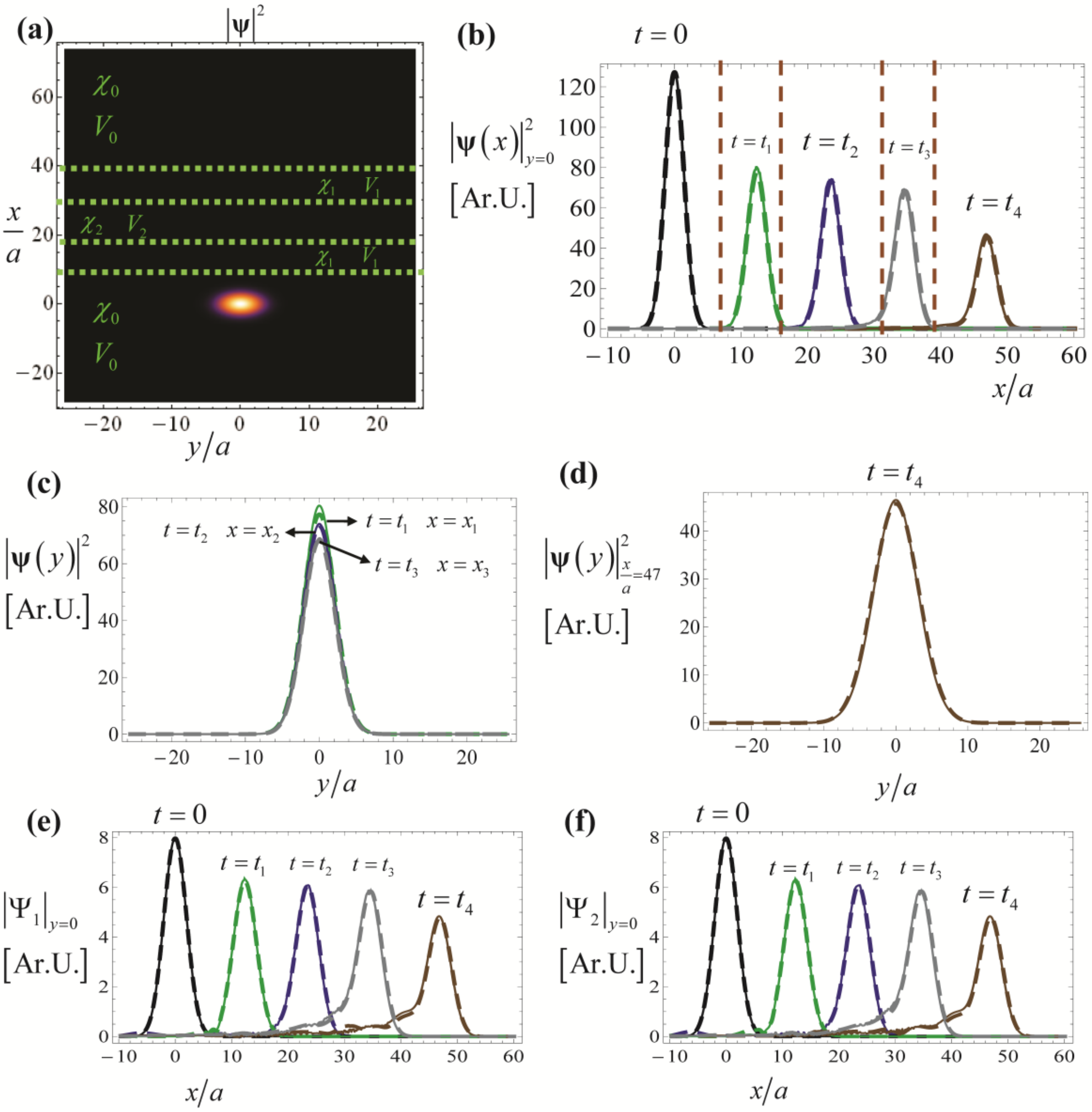

Fig. 10 (color online) **(a)** Geometry of a graphene heterostructure formed by superlattices with anisotropy ratios $\chi_1$ and $\chi_2$ and average potentials $V_1$ and $V_2$ embedded in pristine graphene ( $\chi_0 = 1$ and $V_0 = 0$ ). **(b)** Longitudinal profile of $|\psi|^2$ sampled at $t = 0$ (black curves), $t_1 = 2000\Delta_t$ (green curves), $t_2 = 3800\Delta_t$ (blue curves), $t_3 = 5600\Delta_t$ (gray curves) and $t_4 = 7600\Delta_t$ (brown curves). **(c)** Transverse profile of $|\psi|^2$ sampled at the same time instants as in **(b)**. **(d)** Similar to **(c)** but the sampling time is $t = t_4$. **(e)** Longitudinal profile of $\Psi_1$ sampled at the same time instants as in **(b). (f)** Similar to **(e)** but for $\Psi_2$. In all the examples, the results calculated with the exact microscopic theory are depicted with dashed lines and the continuum approximation results are depicted with solid thick lines.

To highlight the agreement between both models, we also calculated the transverse profiles of the wave function at the same time instants. As seen in Figs. 10**c)** and 10**d**), there is an almost exact agreement between the two formalisms, and only in the case $t = t_1$, the profiles are not exactly overlapped. Figures 10**e**) and 10**f**) also reveal that the agreement is similarly good for the individual components of the pseudo-spinor. Interestingly, one may detect in these profiles the presence of some trembling motion in the wave function. Such a feature is perfectly captured by the continuum approximation. The trembling motion trailing the main beam is plausibly a manifestation of the Zitterbewegung effect [47-51], caused by the interference between positive and negative energy states. In our problem, the Zitterbewegung is due to the fact that states with energies $E - V < 0$ may be excited in the graphene superlattice regions and hence coexist with states with $E - V > 0$. Indeed, the presence of interfaces may imply a redistribution of the relative weight of the waves with $E - V > 0$ (forward waves) and the waves $E - V < 0$ (backward waves, with opposite phase direction). Note that due to the Klein tunneling [3-5] the reflections are rather weak in our problem, and hence the pulse broadening should not be attributed to them.

## VI. Conclusion

In this article it was numerically demonstrated that an effective Hamiltonian proposed in an earlier work (Ref. [27]) can be used to characterize the electron transport and the time evolution of a given initial state in complex graphene heterostructures. To this end, a FDTD algorithm was developed to study the electron wave propagation in the frame of the effective medium theory. We derived the conditions of stability of the algorithm and proposed a way to mimic open boundaries based on an absorbing boundary condition. Importantly, the numerical results reveal that the effective Hamiltonian accurately describes both the extended stationary states (continuous spectrum) and the time dynamics of an initial wave packet, provided the initial state is not more localized than the characteristic spatial period of the superlattice. Surprisingly, this property holds even for strong modulations of the electric potential that lead to the emergence of new Dirac points (e.g., for $V_{osc} a / \hbar v_F \approx 7.55$ [12]). Moreover, the numerical results confirm that within the effective Hamiltonian framework the correct boundary condition does not correspond to the continuity of the pseudospinor. We envision that the developed computational tools can be useful to characterize complex graphene electronic devices and may be generalized to determine the optical response (e.g. the conductivity) of graphene superlattices.

## Appendix A: The Hamiltonian of graphene heterostructures

As discussed in the main text, a graphene superlattice may be regarded as a continuum characterized by the Hamiltonian (2). This Hamiltonian can be used to describe the electron wave propagation in a homogeneous unbounded region. Crucially, it was demonstrated in Ref. [29] that for graphene-based heterostructures, e.g. at an interface between a graphene superlattice (regarded as a continuum) and pristine graphene, the boundary conditions at the interface are not trivial and *do not* imply the continuity of the pseudospinor. Indeed, it is possible to ensure the continuity of the probability density current $j_x = v_F \boldsymbol{\psi}^* \cdot \sigma_x \cdot \boldsymbol{\psi}$ at an interface ($x = x_0$) between two distinct graphene based nanomaterials without having a continuous pseudospinor [29]. In particular in Ref. [29], a generalized boundary condition of the type $\mathbf{U}_{x_0^-} \cdot \boldsymbol{\psi}\left(x_0^-\right) = \mathbf{U}_{x_0^+} \cdot \boldsymbol{\psi}\left(x_0^+\right)$ was put forward to characterize a step-like discontinuity of the anisotropy ratio parameter, where $\mathbf{U}$ is a unitary operator of the form $\mathbf{U} = e^{iu\boldsymbol{\sigma}_x} = \begin{pmatrix} \cos u & i\sin u \\ i\sin u & \cos u \end{pmatrix}$, with $u(\chi)$ some function that depends on the microscopic potential. Within this theory, it is clear that the term $-i\hbar v_F \boldsymbol{\sigma}_x \frac{\partial}{\partial x}$ in Eq. (2) needs to be replaced by $-i\hbar v_F \boldsymbol{\sigma}_x \mathbf{U}^{-1} \frac{\partial}{\partial x} \mathbf{U}$. Hence, we should write the Schrödinger equation as:

$$i\hbar \frac{\partial}{\partial t} \boldsymbol{\psi} = -i\hbar v_F \left( \boldsymbol{\sigma}_x \mathbf{U}^{-1} \frac{\partial}{\partial x} \mathbf{U} + \chi \boldsymbol{\sigma}_y \frac{\partial}{\partial y} \right) \boldsymbol{\psi} + V_{av} \boldsymbol{\psi} . \tag{A1}$$

Importantly, this modified equation holds at all points of space, including the interfaces. Using the formulas $\mathbf{U}^{-1} \frac{\partial}{\partial x} \mathbf{U} = \frac{\partial}{\partial x} + \left( \mathbf{U}^{-1} \cdot \frac{\partial \mathbf{U}}{\partial x} \right) = \frac{\partial}{\partial x} + \frac{\partial u}{dx} \left( \mathbf{U}^{-1} \cdot \frac{d\mathbf{U}}{du} \right)$ and $\mathbf{U}^{-1} \cdot \frac{d\mathbf{U}}{du} = i\boldsymbol{\sigma}_x$, we may simplify Eq. (A1) as follows:

$$i\hbar \frac{\partial}{\partial t} \boldsymbol{\psi} = -i\hbar v_F \left( \boldsymbol{\sigma}_x \frac{\partial}{\partial x} + \chi \boldsymbol{\sigma}_y \frac{\partial}{\partial y} \right) \boldsymbol{\psi} + \left( V_{av} + \frac{\partial u}{\partial x} \hbar v_F \right) \boldsymbol{\psi} . \tag{A2}$$

The above equation is consistent with the effective Hamiltonian in Eq. (5) of the main text. The function $u(\chi)$ may be heuristically determined with a numerical fitting as explained in Ref. [29]. In case the microscopic electrostatic potential is characterized by a sinusoidal spatial variation (Fig. 1), we found that the function $u$ is given by:

$$u \approx 0.59 \arccos\left(1 + 1.42(\chi - 1)\right) . \tag{A3}$$

## Appendix B: Stability of the FDTD algorithm

Here, it is shown that the update equations (9) are numerically stable provided the time step is sufficiently small. In the following, it is assumed that the structure is spatially homogeneous ($\chi$ and $V_{ef}$ are independent of the spatial coordinates) and that there is no electron source ($\mathbf{j} = 0$). Our aim is to characterize the stationary states of the system. To this end, we look for plane-wave type solutions of Eq. (9) with:

$$\begin{pmatrix} \Psi_{1,p+1,q}^{n} \\ \Psi_{2,p+1/2,q+1/2}^{n+1/2} \end{pmatrix} = \xi_p \begin{pmatrix} \Psi_{1,p,q}^{n} \\ \Psi_{2,p-1/2,q+1/2}^{n+1/2} \end{pmatrix}, \tag{B1a}$$

$$\begin{pmatrix} \Psi_{1,p,q+1}^{n} \\ \Psi_{2,p+1/2,q+1/2}^{n+1/2} \end{pmatrix} = \xi_q \begin{pmatrix} \Psi_{1,p,q}^{n} \\ \Psi_{2,p+1/2,q-1/2}^{n+1/2} \end{pmatrix}, \tag{B1b}$$

where $\xi_p = e^{i\theta_p}$ and $\xi_q = e^{i\theta_q}$ are the spatial phase-shifts between adjacent nodes. In addition, we assume a time variation such that $\Psi_{1,p,q}^{n+1} = \lambda \Psi_{1,p,q}^{n}$ and $\Psi_{2,p,q}^{n+1/2} = \lambda \Psi_{2,p,q}^{n-1/2}$ with $\lambda = \lambda\left(\xi_p, \xi_q\right)$. The algorithm is stable provided $|\lambda| \le 1$ for arbitrary values of the spatial phase-shifts $\xi_p, \xi_q$ with $\left|\xi_p\right| = \left|\xi_q\right| = 1$. Substituting Eq. (B1) into Eq. (9) it is readily found that the plane wave states must satisfy:

$$\lambda \Psi_{1,p,q}^{n}\left(1-\frac{V}{i\hbar 2}\Delta_t\right) = \Psi_{1,p,q}^{n}\left(1+\frac{V}{i\hbar 2}\Delta_t\right) - v_F \Delta_t \lambda \left[\left(\frac{1}{2\Delta_x} - i\frac{\chi}{2\Delta_y}\right) - \left(\frac{1}{2\Delta_x} + i\frac{\chi}{2\Delta_y}\right)\xi_p^{-1} + \left(\frac{1}{2\Delta_x} + i\frac{\chi}{2\Delta_y}\right)\xi_q^{-1} - \left(\frac{1}{2\Delta_x} - i\frac{\chi}{2\Delta_y}\right)\xi_p^{-1}\xi_q^{-1}\right]\Psi_{2,p+\frac{1}{2},q+\frac{1}{2}}^{n-\frac{1}{2}} \tag{B2a}$$

$$\lambda \Psi_{2,p+\frac{1}{2},q+\frac{1}{2}}^{n-\frac{1}{2}}\left(1-\frac{V}{i\hbar 2}\Delta_t\right) = \Psi_{2,p+\frac{1}{2},q+\frac{1}{2}}^{n-1/2}\left(1+\frac{V}{i\hbar 2}\Delta_t\right) - v_F \Delta_t \left[\left(\frac{1}{2\Delta_x} + i\frac{\chi}{2\Delta_y}\right)\xi_p\xi_q - \left(\frac{1}{2\Delta_x} - i\frac{\chi}{2\Delta_y}\right)\xi_q + \left(\frac{1}{2\Delta_x} - i\frac{\chi}{2\Delta_y}\right)\xi_p - \left(\frac{1}{2\Delta_x} + i\frac{\chi}{2\Delta_y}\right)\right]\Psi_{1,p,q}^{n} \tag{B2b}$$

The above system of equations may be written in a compact matrix form as follows:

$$\begin{pmatrix} \frac{1}{\lambda}\left(\lambda - 1 - \frac{V}{i\hbar 2}\Delta_t\left(\lambda+1\right)\right) & v_F \Delta_t D_- \\ v_F \Delta_t D_+ & \left(\lambda - 1 - \frac{V}{i\hbar 2}\Delta_t\left(\lambda+1\right)\right) \end{pmatrix} \begin{pmatrix} \Psi_{1,p,q}^{n} \\ \Psi_{2,p+\frac{1}{2},q+\frac{1}{2}}^{n-\frac{1}{2}} \end{pmatrix} = 0, \tag{B3a}$$

$$D_- = \left(\frac{1}{2\Delta_x} - i\frac{\chi}{2\Delta_y}\right) - \left(\frac{1}{2\Delta_x} + i\frac{\chi}{2\Delta_y}\right)\xi_p^{-1} + \left(\frac{1}{2\Delta_x} + i\frac{\chi}{2\Delta_y}\right)\xi_q^{-1} - \left(\frac{1}{2\Delta_x} - i\frac{\chi}{2\Delta_y}\right)\xi_p^{-1}\xi_q^{-1}, \tag{B3b}$$

$$D_+ = \left(\frac{1}{2\Delta_x} + i\frac{\chi}{2\Delta_y}\right)\xi_p\xi_q - \left(\frac{1}{2\Delta_x} - i\frac{\chi}{2\Delta_y}\right)\xi_q + \left(\frac{1}{2\Delta_x} - i\frac{\chi}{2\Delta_y}\right)\xi_p - \left(\frac{1}{2\Delta_x} + i\frac{\chi}{2\Delta_y}\right). \tag{B3c}$$

To have nontrivial solutions, $\lambda$ is required to satisfy the characteristic equation:

$$\frac{1}{\lambda}\left(\lambda - 1 - \frac{V}{i\hbar 2}\Delta_t\left(\lambda+1\right)\right)^2 - \left(v_F \Delta_t\right)^2 D_- D_+ = 0. \tag{B4}$$

Using $\xi_p = e^{i\theta_p}$ and $\xi_q = e^{i\theta_q}$ it can be shown that:

$$D_- D_+ = \frac{-4\chi^2}{\Delta_x^2 \Delta_y^2} \left[ \Delta_x^2 \cos^2 \frac{\theta_p}{2} \sin^2 \frac{\theta_q}{2} + \frac{\Delta_y^2}{\chi^2} \sin^2 \frac{\theta_p}{2} \cos^2 \frac{\theta_q}{2} \right]. \tag{B5}$$

The solutions of characteristic equation are:

$$\lambda = \frac{1}{(A-i)^2} \left[ -1 - A^2 + \frac{B^2}{2} \pm B \sqrt{\left(\frac{B}{2}\right)^2 - A^2 - 1} \right], \tag{B6}$$

with the parameters $A = \frac{V}{2\hbar}\Delta_t$ and $B^2 = -(v_F \Delta_t)^2 D_- D_+ > 0$ real-valued. It is straightforward to check that if $\left(\frac{B}{2}\right)^2 - A^2 - 1 < 0$ then

$$|\lambda| = \frac{1}{(1+A^2)} \left[ \left(1 + A^2 - \frac{B^2}{2}\right)^2 + B^2 \left(1 + A^2 - \frac{B^2}{4}\right) \right]^{1/2} = 1. \tag{B7}$$

Thus, we can conclude that the FDTD algorithm is stable when $\left(\frac{B}{2}\right)^2 - A^2 - 1 < 0$. This condition is equivalent to

$$\frac{\chi^2}{\Delta_x^2 \Delta_y^2} \left[ \Delta_x^2 \cos^2 \frac{\theta_p}{2} \sin^2 \frac{\theta_q}{2} + \frac{\Delta_y^2}{\chi^2} \sin^2 \frac{\theta_p}{2} \cos^2 \frac{\theta_q}{2} \right] (v_F \Delta_t)^2 < 1 + \left(\frac{V}{2\hbar}\Delta_t\right)^2. \tag{B8}$$

The above inequality should hold for arbitrary values of the spatial phase-shifts $\xi_p = e^{i\theta_p}$ and $\xi_q = e^{i\theta_q}$. To satisfy this stability criterion it is sufficient to ensure that $\frac{\chi^2}{\Delta_x^2 \Delta_y^2}\left(\Delta_x^2 + \frac{\Delta_y^2}{\chi^2}\right)(v_F \Delta_t)^2 < 1$, which is the same as Eq. (10).

## Appendix C: The perfectly matched layer

Here, it is explained how an unbounded graphene region can be mimicked using the FDTD method. This can be done with a "perfectly matched layer" (PML) that ideally should "absorb" any incoming electron wave. We propose to do this by tailoring the electrostatic potential $V$ such that in the PML region $V$ is complex-valued and the wave propagation is damped. Analogous ideas are used in electromagnetics to imitate the wave propagation in unbounded regions [52]. It should be noted that a Hamiltonian with a complex-valued $V$ is not Hermitian and hence it is compatible with damping in the wave propagation. It can be easily verified that to have a damped wave propagation the electric potential $V$ must have a negative imaginary part: $\text{Im}\{V\} < 0$. For example, for pristine graphene the relevant Hamiltonian is $\hat{H} = -i\hbar v_F \boldsymbol{\sigma} \cdot \nabla + V$ and hence from the Schrödinger equation one obtains a modified probability current conservation law $\nabla \cdot \mathbf{j}_e + \frac{\partial W}{\partial t} + p_{loss} = 0$, being $\mathbf{j}_e = v_F \left( \boldsymbol{\psi}^* \cdot \boldsymbol{\sigma}_x \cdot \boldsymbol{\psi} \hat{\mathbf{x}} + \boldsymbol{\psi}^* \cdot \boldsymbol{\sigma}_y \cdot \boldsymbol{\psi} \hat{\mathbf{y}} \right)$ the probability current, $W = |\boldsymbol{\psi}|^2$ the probability density and $p_{loss} = -\text{Im}\{V\} \frac{2}{\hbar} |\boldsymbol{\psi}|^2$. The term $p_{loss}$ is the instantaneous rate of "probability absorption" per unit of volume, and hence should be positive to have a damped propagation. Thus, it is required that $\text{Im}\{V\} < 0$.

In our numerical implementation the PML completely surrounds the material region of interest with dimensions $L_x \times L_y$ (Fig. 11). In a time evolution problem of a certain initial localized electronic state, the wave will eventually reach and be absorbed by the PML. Hence, after a sufficiently long time it will eventually abandon the region of interest (confined by the PML boundary) so that $\int |\boldsymbol{\psi}|^2 \, dxdy \to 0$. The thicknesses of the vertical and horizontal PML regions are $d_{PML,x}$ and $d_{PML,y}$, respectively (Fig. 11).

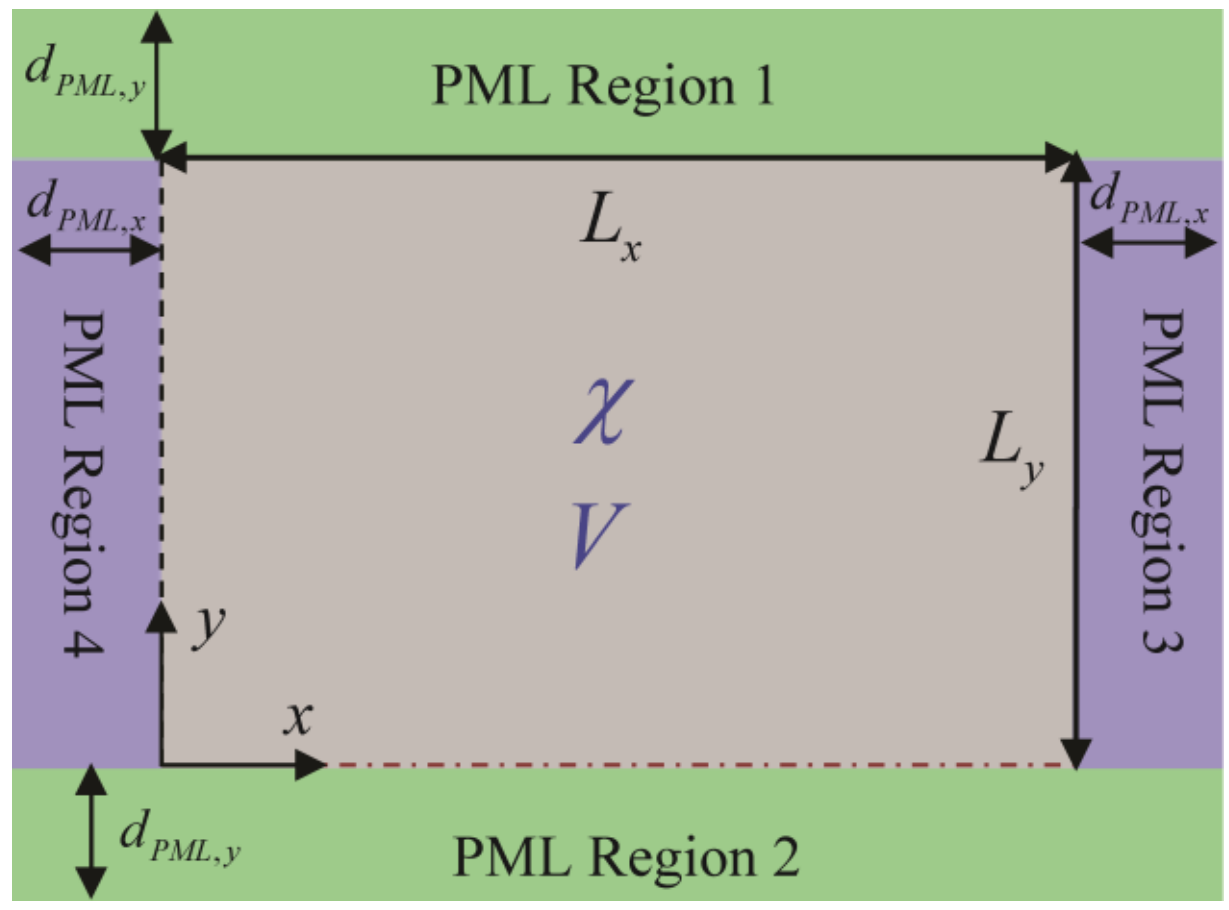

Fig. 11 (color online) Illustration of the PML regions surrounding the relevant computational domain, i.e. a graphene heterostructure with spatially varying anisotropy ratio $\chi$ and potential $V$ .

In our numerical implementation it is assumed that $\chi$ in the PML regions is the same as in the material adjacent to the PML. Similarly, the real part of the potential $V$ is the same as for the neighboring material ($V_0$). We adopted the following potential distribution in the regions 1 and 2 of the PML:

$$V - V_0 = -i0.7\hbar v_F \left( \frac{2\pi}{d_{PML,y}} \frac{d_y}{d_{PML,y}} \right). \tag{C1}$$

where $0 < d_y < d_{y,PML}$ is the distance to the PML boundary (e.g. for $-d_{PML,y} < y < 0$ one has $d_y = |y|$ and for $L_y < y < L_y + d_{PML,y}$ one has $d_y = y - L_y$ ). Note that $\text{Im}\{V\} < 0$ to ensure the wave absorption in the PML. Similarly, in regions 2 and 3 of the PML we use:

$$V - V_0 = -i0.7\hbar v_F \left( \frac{2\pi}{d_{PML,x}} \frac{d_x}{d_{PML,x}} \right), \tag{C2}$$

where $0 < d_x < d_{x,PML}$ is the distance to the relevant PML boundary. Note that in regions 2 and 3 the imaginary part of the potential is continuously increased (in absolute value) until it reaches a constant value. This ensures a good absorption of the electron waves. In our simulations we used a PML with thicknesses $d_{PML,x} = 15a$ and $d_{PML,y} = 15a$ .

# Appendix D: The current source

As discussed in the main text, it is possible to study the extended electronic states (the continuous spectrum) with a given energy $E = E_0$ (e.g. the scattering of a plane wave by a graphene heterostructure) using the FDTD method. More specifically, an incoming incident wave can be emulated by introducing a fictitious electron source $\mathbf{j} = \begin{pmatrix} j_1 \\ j_2 \end{pmatrix}$ in the massless Dirac equation (4). For plane wave incidence, the following current distribution is adopted,

$$\mathbf{j} = \mathbf{j}_0 e^{ik_{y0}(y-y_0)} e^{-i\omega_0 t} \delta(x - x_0), \qquad t > 0, \tag{D1}$$

where $\omega_0 = E_0 / \hbar$, $\mathbf{r}_0 = (x_0, y_0)$ defines the location of the source, and $\mathbf{j}_0$ is the vector

$$\mathbf{j}_0 = \boldsymbol{\sigma}_x \cdot \begin{pmatrix} 1 \\ \dfrac{\hbar v_F}{E_0 - V}\left(k_{x0} + i\chi k_{y0}\right) \end{pmatrix}. \tag{D2}$$

In the above $\mathbf{k}_0 = (k_{x0}, k_{y0})$ is the wave vector associated with the plane wave, which satisfies $|E_0 - V| = \hbar v_F \sqrt{k_{x0}^2 + \chi^2 k_{y0}^2}$. If $E_0 > V$ one should choose $k_{x0} > 0$, otherwise $k_{x0} < 0$.

Next it is shown that in a steady-state ($t \to \infty$) this source emits a plane wave of the form $\boldsymbol{\psi} = \begin{pmatrix} 1 \\ \dfrac{\hbar v_F}{E_0 - V}\left(k_{x0} + i\chi k_{y0}\right) \end{pmatrix} e^{i\mathbf{k}_0 \cdot (\mathbf{r} - \mathbf{r}_0)}$ into the region $x > x_0$. Indeed, in a steady-state ($t \to \infty$) the solution of (4) under the excitation (D1) satisfies:

$$\boldsymbol{\psi} = \begin{cases} C_+ \begin{pmatrix} 1 \\ \dfrac{\hbar v_F}{E_0 - V}\left(k_{x0} + i\chi k_{y0}\right) \end{pmatrix} e^{ik_{x0}(x-x_0)} e^{ik_{y0}(y-y_0)} e^{-i\omega_0 t}, & x > x_0 \\ C_- \begin{pmatrix} 1 \\ \dfrac{\hbar v_F}{E_0 - V}\left(-k_{x0} + i\chi k_{y0}\right) \end{pmatrix} e^{-ik_{x0}(x-x_0)} e^{ik_{y0}(y-y_0)} e^{-i\omega_0 t}, & x < x_0 \end{cases} \tag{D3}$$

where $C_\pm$ are unknown constants. From here it follows that at $x = x_0$ the term $-i\hbar v_F \sigma_x \frac{\partial \boldsymbol{\psi}}{\partial x}$ in Eq. (4) originates a $\delta$-type singularity that should cancel out the term $+i\hbar v_F \mathbf{j}$. This is only possible if the vector $\mathbf{j}_0$ satisfies:

$$\mathbf{j}_0 = C_+ \boldsymbol{\sigma}_x \cdot \begin{pmatrix} 1 \\ \frac{\hbar v_F}{E_0 - V}\left(k_{x0} + i\chi k_{y0}\right) \end{pmatrix} - C_- \boldsymbol{\sigma}_x \cdot \begin{pmatrix} 1 \\ \frac{\hbar v_F}{E_0 - V}\left(-k_{x0} + i\chi k_{y0}\right) \end{pmatrix}. \tag{D4}$$

Hence, in order that in the stationary state all the "electrons" are launched into the semi-space $x > x_0$ one should have $C_+ = 1, C_- = 0$, which yields (D2).

An incoming Gaussian beam with radius $R_G$ may also be generated with an excitation of the form (D1) using a Gaussian modulating term:

$$\mathbf{j}_0 = \boldsymbol{\sigma}_x \cdot \begin{pmatrix} 1 \\ \frac{\hbar v_F}{E_0 - V}\left(k_{x0} + i\chi k_{y0}\right) \end{pmatrix} e^{-\frac{(y-y_0)^2}{R_G^2}}. \tag{D5}$$

Note that in the numerical implementation $\delta\left(x - x_0\right) \to \frac{1}{\Delta_x}\delta_{p,p_0}$ wherein $p_0$ determines the grid line wherein the fictitious source is placed.

## Acknowledgements

This work was funded by Fundação para Ciência e a Tecnologia under projects PTDC/EEI-TEL/2764/2012 and UID/EEA/50008/2013. D. E. Fernandes acknowledges support by Fundação para a Ciência e a Tecnologia, Programa Operacional Potencial Humano/POPH, and the cofinancing of Fundo Social Europeu under the fellowship SFRH/BD/70893/2010.